# Inclusive learning for quantum computing: supporting the aims of quantum literacy using the puzzle game Quantum Odyssey

**Authors**

**Laurentiu Nita, Nicholas Chancellor, Laura Mazzoli Smith, Helen Cramman, Gulsah Dost**

**Abstract**

With a vast domain of applications and now having quantum computing hardware available for commercial use, an education challenge arises in getting people of various background to become quantum literate. Quantum Odyssey is a new piece of computer software that promises to be a medium where people can learn quantum computing without any previous requirements. It aims to achieve this through visual cues and puzzle play, without requiring the user to possess a background in computer coding or even linear algebra, which are traditionally a must to work on quantum algorithms. In this paper we report our findings on an UKRI (United Kingdom Research and Innovation) Citizen Science grant that involves using Quantum Odyssey to teach how to construct quantum computing algorithms. Sessions involved 30 minutes of play, with 10 groups of 5 students, ranging between 11 to 18 years old, in two schools in the UK. Results show the Quantum Odyssey visual methods are efficient in portraying counter-intuitive quantum computational logic in a visual and interactive form. This enabled untrained participants to quickly grasp difficult concepts in an intuitive way and solve problems that are traditionally given today in Masters level courses in a mathematical form. The results also show an increased interest in quantum physics after play, a higher openness and curiosity to learn the mathematics behind computing on quantum systems. Participants developed a visual, rather than mathematical intuition, that enabled them to understand and correctly answer entry level technical quantum information science.

**Introduction**

Significant advances have recently been made in the development of quantum computers, moving to a stage where real applications are in sight (Kandala 2017). The potential ability of quantum computers to perform calculations of a size and complexity which is not possible with classical computing power will have substantial benefits for fields such as machine learning, artificial intelligence, materials chemistry and pharmaceuticals. However, it also brings significant risks, with the computational power required to pose a significant threat to cybersecurity and encryption systems such as those used for banking, health and government data (Shor 1997). Therefore, to maximise the benefits of quantum computing and to reduce the risks, there is a need to upskill the current workforce and begin the training of future programmers and decision makers in the counter-intuitive ways of quantum computational thinking. The concept of quantum literacy (Nita et al., 2020) has been developed to formalize this need as an educational challenge.

In this paper we report on a UKRI Citizen Science grant[1] in which we carried out trials with students on the potential of a puzzle game, introduced in Nita et al (2020), to support non-specialists in quantum mechanics to develop their understanding of quantum computational thinking. Participants in this study were viewed very much as collaborators in the development of this innovative puzzle game. It was also a digital innovation pilot in the sense that the project focused specifically on testing or extending the functionality of the existing puzzle game. Future applications of quantum computing will be in areas in which there is no current expertise or understanding of this field, therefore, involvement of participants from the early stages of development of an innovative tool such as this will maximise the future learning benefits for users.

---

[1] Funded by UKRI (United Kingdom Research and Innovation) grant number BB/T018666/1.

The puzzle game used in this project is called Quantum Odyssey and was developed by Quarks Interactive[2]. The ultimate aim of the puzzle game is to enable users to develop an intuitive but still rigorous understanding of the fundamentals of quantum computation, as well as the potential to discover increasingly complex and new quantum algorithms. Because high level tools for quantum computing are not yet fully developed, understanding the underlying building blocks is crucial, unlike in classical computing, where much of the low-level behaviour of the computer can be abstracted away. The principle behind the game design is that the dynamics of classically counter-intuitive processes such as phase amplification can be understood in such a way that even if the mathematics is hard to grasp, they can be intuitively understood by engaging with the visual tool and solving puzzles.

Quantum Odyssey consists of a visual Plinko (grid of pins) board system where coloured balls travel down the board following different tracks. Players are tasked to solve increasingly complex puzzles using picture tiles to change the path the balls take, or to introduce actions on the balls. Behind the scenes, each picture tile accurately represents a quantum mechanical rule and the player's sequence of tiles is writing a functioning quantum algorithm. From this visualisation tool, the aim is that the players will be able to learn about fundamental principles behind quantum mechanics such as superposition and interference. However, crucially, the player should not need to have any knowledge of the complex mathematics behind the graphical interface to be able to successfully complete the puzzles. The aim is that the puzzles therefore enable the player to develop an intuition about quantum mechanics, as well as generating genuine quantum computing algorithms, which can be implemented on quantum computers. A fuller description along with visual representations of the game is available in Nita et al. (2020).

**Game-based learning and gamification**

Quantum Odyssey is a puzzle-game and therefore we are exploring game-based learning, that is the exposure of students to games that have an educational objective to be achieved through the game play (Kim, Park and Baek, 2009). We would suggest that Quantum Odyssey is also usefully understood through exploring some elements of gamification, where learning utilises gaming elements so that it becomes more engaging and challenging for students (Deterding et al., 2011; Landers et al., 2018) because of the nature of the accurate representation of quantum states through the game (described further below and in Nita et al 2020). Other factors supporting the use of literature on game-based learning and gamification includes the development of scaffolded instruction, which is dependent on individual student's needs (Hanus and Fox, 2015); and the rapid and continuous feedback afforded by games (Kapp, 2012a), that allows students the opportunity to fail without fear when learning (Lee and Hamer, 2011). An unexpected consequence of gamification has been that teachers find their role shifted from lecturer to guide because students control their own learning (Rose, 2015), which we would argue is the potential of Quantum Odyssey as well. In the context of STEM education, this type of instructional method is very well suited to creating an environment that subverts the conventional classroom by providing a space to encourage exploration and making connections. Nevertheless, there is still a shortage of data and studies on STEM games and gamification (Ludwig et al., 2020) that this paper works to counteract.

---

[2] Quantum Odyssey was provided under licence from Quarks Interactive Ltd.

Rojas et al (2016) presented an initial overview of the research literature that is available in terms of gamification within STEM fields in Higher Education and showed that Computer Science (CS) is the dominant STEM field being studied; a combination of points, badges and leaderboards are the most commonly used tools; and most studies view student engagement as the key dependent variable. Similarly, Dichev and Dicheva (2017) also highlighted that the vast majority of gamification studies focus on CS and Information Technology (IT) instruction.

Studies in schools have shown mixed findings relating to the impact of gamification. While some studies have demonstrated no effect from the use of interactive materials compared with alternative methods on student performance (Reinhold, 2019), other studies show statistically significant improvements when using STEM games in schools (Freina et al., 2018). Research has shown gamification mechanisms assist the teaching of STEM subjects by making them engaging for students (Lynch et al., 2018). Kusuma et al. (2018) reviewed 33 papers relating to gamification models in education as applied in four domain applications: language, history, STEM and generic. The findings from the review show some representative gamification models that could be used to increase motivation and maximize achievement and engagement in learning activities. August et al. (2015) established a Virtual Engineering Sciences Learning Lab (VESLL) to support STEM education and exposed students to various quantitative skills and concepts through visualization, collaborative games, and problem solving with realistic learning activities. The results demonstrated high levels of student interest in VESLL's potential as a supplementary instructional tool, and improved learning experiences.

Gamification can indirectly prompt the student to acquire more knowledge and skills due to the effect on engagement and motivation (Huang and Soman, 2013). Rose (2015) evaluated the potential for the gamification of on-line undergraduate physics content as a mechanism to enhance student learning and improve learner motivation. Rasool et al. (2014) focused on gamifying in an adaptive learning environment for physics problem solving, utilizing Adaptive Learning Environment for Problem Solving (ALEPS). ALEPS is based on Polya's (1945) problem solving strategies, which include, understanding, planning, implementing and checking and using visualization to highlight and explicate schemata. Similarly, dela Cruz, Tolentino and Roleda (2020) evaluated the effects of gamification on the motivation of freshmen engineering students studying physics. To determine how the elements of video games can be used to enhance students' motivation to learn about physics, Cruz and Roleda (2018) implemented gamification in two grade 10 classes over a month. Through the use of journal entries, interviews and surveys, the results revealed that students had a significant improvement in their test performance following the intervention. In addition, the students were shown to be more motivated to learn the subject, because of the enjoyment of the fun atmosphere introduced by the gaming elements. All four studies' results showed that gaming techniques, systems and instructions were significantly correlated to students' motivation and engagement with physics, namely: intrinsic motivation, self-efficacy, grade motivation, career motivation and self-determination.

**Visualisation in STEM education**

Visualization is central to learning, especially in the sciences, where students have to navigate within and between modes of representation, as argued by Gilbert (2005). This led Gilbert (2005) to argue that students, especially science students, integrate a private mental model with the consensus – and hence scientific – models available. Visualizations have been shown to help students develop scientific concepts (Honey and Hilton, 2011) and student achievement in science is supported by access to multi-media modes of representation (Ardac and Akaygun, 2004). This is largely influenced by Piaget and Inhelder's (1967) age-based developmental phases, where they demonstrate that at around 12 years of age a child becomes able to visualize concepts such as area, volume, and distance in combination with translation, rotation, and reflection, all of which are essential for learning physics. Wiley (1990) also presented a hierarchy of visual learning, partially based on Piaget's work and comprising of three stages of visual learning; i.e. visual cognition, visual production and visual resolve. A person who has the ability to solve a physics problem more readily is often described as having physical intuition (Werner, Becker and Claudete, 2016). To develop physical intuition is not easy, but it can be assisted by visualizing and gamifying problematic phenomena. The visualization and gamification of a phenomenon can mitigate its abstract nature (Ferreira, Filho and Ferreira, 2019) making it easier to solve (Suyatna et al., 2017).

There is an interesting and relevant debate in how far the objects of science are aided by the epistemological status of mental models. For instance, Gilbert (2005) refers to the different status of a private mental model, the representation made visually and internally by an individual through a constructivist process within their own particular meaning frame (von Glaserfeld 1995). The consensus, or scientific model, however, is one that is widely endorsed to be agreed upon and epistemologically largely thought to represent structures of a realist world as accurately as possible. Karakostas and Hadsidaki (2005) question what it means to suggest that the objects of science can constitute personal constructions in a social constructivist model of visualisation, common in social constructivist theories of learning (e.g von Glaserfeld, 1995). They add to this the paradigmatic complexity in learning quantum physics, when students are grounded in a Newtonian ontology with a clear demarcation of knowing subject and object of knowledge in the world as well as 'the absence of genuine indeterminism in the course of events or of an element of chance in the measurement process' (2005; 1704). They suggest, however, that standard quantum mechanics;

> 'systematically violates the conception of separability…The generic phenomenon of quantum nonseparability, experimentally confirmed for the first time in the early 1980s, precludes in a novel way the possibility of defining individual objects independently of the conditions under which their behaviour is manifested'. (Karakostas and Hadsidaki 2005)

These issues, particularly that of methodological holism, where the functioning of the physical world cannot be reduced to its constituent parts as in quantum mechanics, necessarily impacts on any consideration of visualisation to aid learning in quantum computation. If 'quantum mechanical formalism seems only to allow a detailed description of reality that is co-determined by the specification of a measurement context' (Karakostas and Hadzidaki, 2005: 622) then visualization of any aspects of such reality in learning could be aided by a contextual realism that Karakostas (2004) terms 'active scientific realism', acknowledging the active role of the participatory subject in a reality of rule-bound objects. We take from this work the need for there to be visual models that strongly scaffold the students' own mental models and support the need for conceptual shifts from the objective realism of classical Newtonian physics, to the methodological holism of quantum mechanics. Arguably a powerful visualization tool that is faithful to this contextual realism and the complex underlying scientific phenomena is of great value here, alongside one that can help students to 'manage uncertainty'. This is important for scientific literacy in general (Britt et al 2014; Kienhues et al. 2020), but arguably all the more so given the additional uncertainty inherent in the epistemic shift from classical to the quantum reality.

One area relevant to our quantum computation puzzle game, in which research has been carried out to investigate the use of visualisation to support students in understanding complex algorithms, is in the use of the RC4 algorithm, central to cryptography (the field of science applied to secure data and information). Sriadhi et al (2018) provide a thorough understanding of the workings of the RC4 cryptographic algorithms. Their paper focuses on tools employing visualization to demonstrate the work of the RC4 algorithm to facilitate learners' understanding of cryptography. Ryoo and Linn (2014) evaluated guidance to support students, when interacting with the dynamic visualizations associated with complex scientific phenomena in instructional inquiry. Their findings reported that well-designed guidance and pair work enables students to generate valuable explanations. This in turn suggests promising designs for online instruction, featuring dynamic visualizations.

**Games and visualization in quantum computing education**

It is difficult to teach the concepts of quantum computing via traditional didactic teaching methods, due both to epistemological challenges discussed above and the lack of availability of specialized quantum computing systems and related equipment, which complicates tactile learning and experimentation. Due to its abstract nature and the often counterintuitive results, simulations and visualisations might arguably be ideal for teaching quantum mechanics (Kohnle et al., 2012). Research-based interactive simulations for quantum mechanics have been shown to improve students' understanding (Kohnle et al., 2013). In addition, a number of tools have been developed for quantum visualization, including a wide array of simulations, including PhET (McKagan et al., 2008), QuVIS (Kohnle et al, 2012) and QuILT (Singh, 2008). For example, the QuVis Quantum Mechanics Visualization project provides interactive simulations to teach quantum mechanics (Kohnle et al., 2010). Each animation takes students through a concept systematically in steps. Each step explains one aspect of the animation in detail, through accompanying text, graphs, and highlights.

Kohnle, Baily and Ruby (2014) described efforts in QuVis simulations to refine visual representations of a single-photon superposition state. Their outcomes described the incorporation of a revised visualization of all QuVis single-photon simulations. The class used a revised visualization showing a lower frequency of incorrect ideas regarding quantum superposition, such as the photon splitting into two half-energy components. Kohnle et al. (2015) described the evaluation of simulations focusing on two-level systems, which were developed as part of the Institute of Physics, Quantum Physics resources. They offer evidence that these simulations help students learn quantum mechanics concepts at introductory and advanced undergraduate level.

The game, Quantum Moves (Ornes, 2018), is based on a real world problem from quantum computing. It asks how fast a laser can move an atom between wells in an egg-box-like structure without altering the energy of the atom, which is in a delicate quantum state. In the quantum world, speed and energy are constrained by Heisenberg's uncertainty principle. The aim of the game is to locate the point at which the transition from one place to another is as fast as possible without disturbing the quantum state. The principle of Quantum Moves is to manipulate the potential landscape (Lieberoth et al, 2014). The players manipulate a physics problem though a game interface to solve abstract challenges, which would otherwise be highly intractable.

The Quantum Odyssey puzzle game differs from Quantum Moves in that it is a more abstract representation. While Quantum Moves represents a quantum fluid (a Bose-Einstein condensate) moving in real space, our tool uses an abstract solution space, representing all possible configurations simultaneously. Quantum Moves allows understanding and visualization of important quantum control problems, but is limited in the complexity of entangled states it can represent due to the nature of the representation. At the cost of only being able to represent smaller systems, our tool can represent arbitrary qubit configurations, and therefore allows understanding of more complex many body phenomena which are not directly accessible in quantum fluids. Both types of understanding are valuable and indeed complementary, but the understanding gained from Quantum Moves is not going to directly lend itself to complete understanding of quantum computing. Quantum Odyssey is designed to both visually represent and allow interaction with everything that can be done on small quantum computing systems. The puzzle creator has the tools to enforce various rules and winning conditions that can encode complex and varied problems that are relevant for the creator under the form of puzzles sets, where the challenge for the player is to work out actual quantum computing algorithms that resolve the puzzle set made by the creator. Thus, the only limits on the types of quantum computing problems that can be encoded in Quantum Odyssey come from the size of the visualisation, limited by the developer to systems of up to 5 quantum bits.

Puzzle games range on a spectrum from concrete to abstract. A concrete puzzle game makes use of things players already have strong and accurate intuitions for, and the challenge is simply to combine these elements skilfully (Leifer, 2017). Dorland et al. (2019) developed Save Schrödinger's Cat, a virtual reality puzzle game comprising a classical physics mode as well as a quantum physics mode. The game encourages players to explore the differences between quantum and classical physics in a challenging way. A preliminary evaluation by the developer showed players could effectively identify the various distinguishing features in either mode.

In order to assess transitions in student thinking, Passante and Kohnle (2019) developed a framework to characterize student responses in terms of real and complex mathematical reasoning and classical and quantum visual reasoning. The results indicated that simulation tutorials support the development of visual understanding of time dependence. Specifically, that visual reasoning correlates with improved student performance in the areas of time evolution of wave function and probability density. Chhabra and Das (2016) presented a qualitative analysis of the data obtained when administering a questionnaire to a group of ten undergraduate-level students. The questionnaire contained four visualization-based questions on the topic of 'wave function', which forms the basis for understanding the behaviour of physical systems. The outcome of the study reveals key areas in students' conceptualization of quantum mechanics, including important misconceptions and the observation of a lack of links made by students between quantum mechanics and classical mechanics. It also provides a plausible route by which to address issues at the pedagogical level within the classroom.

A study of undergraduate students using a combined simulation-tutorial within a core quantum mechanics course at two institutions in the UK and USA, found many students do not understand the distinction between solutions for time-independent and time-dependent Schrödinger equations (Passante and Kohnle, 2019). In order to probe various aspects of students' understanding of some of the core ideas associated with quantum mechanics, and especially to establish how they develop within an undergraduate curriculum, Cataloglu and Robinett (2002) developed an assessment instrument designed to test conceptual and visualization understanding of quantum theory. The Quantum Mechanics Visualization Instrument (QMVI) assessment tool, focuses on students' conceptual and visual understanding. Their findings suggest that students struggle in general when they need to connect conceptual ideas with quantitative interpretations. Another notable addition to visualisation tools for educational purposes in quantum physics is the Quantum Composer introduced by Ahmed et al. (2020). This allows the user to build, expand, or explore quantum mechanical simulations by interacting with graphically connectable nodes, each corresponding to a physical concept, mathematical operation, or visualization.

Bøe, Henriksen and Angell (2018) studied an innovative approach implementing features in quantum physics and found that students struggled when the new approach held implicit expectations that varied greatly from how students typically "do physics" in traditional classrooms. There was a need for better alignment between learning activities and learning goals in innovation so that students know what "doing physics" successfully entails. Bjaelde, Pedersen and Sherson (2014) demonstrated how a gamified teaching setup can support student learning during a quantum mechanics course. They found the students that relatively learned the most were those who scored lowest on the pre-test. However, students with high grades also tended to learn more than average students. They highlighted that these findings appear to contradict one another, and further research is required in this area. Kohnle et al. (2010) described animations and animated visualizations for introductory and intermediate-level quantum mechanics instruction. The questionnaires revealed that students on the whole are very positive about interactive visualizations and make substantial use of them.

Early evidence therefore shows that simulations and visualizations of quantum physics experiments and concepts are effective means of conveying the curriculum (Kontogeorgiou, Bellou and Mikropoulos, 2008). For example, Sakurai, and Napolitano (2014) built a 3D simulation of the Stern-Gerlach experiment (SGE), which is a central topic in any introductory courses in quantum physics. The visualization of the experiment illustrates the counterintuitive quantum effects that appear when a spin is measured along orthogonal axes. In addition to aiding learning, research is needed into the fact that quantum games can also directly benefit the science they are based upon, e.g., Decodoku games from Wootton (2017), which allow players to play with and eventually discover better quantum error correction algorithms.

**Description of Quantum Odyssey**

In Quantum Odyssey the players see in real time actual quantum computation fully encoded to visual cues, such as changing in direction, size, number, or colour, the falling balls that navigate a quantum circuit built by the student using puzzle pieces that are quantum gates, and not by performing traditional calculus. The players had to experiment with each quantum gate to see what it does and remember the visual cues and their effects. Although the game came without narration or explanations at this early stage of piloting, each new concept was introduced one at a time to give the players time to experiment, while later puzzles mixed in all the key concepts.

Players interact with the quantum circuit via an interface that consisted of a Circuit Configuration panel (upper part of Figure 1). In the panel, the player was given access to up to three quantum gates (H, X and CTRL) that could be dragged and dropped in the Quantum Circuit Panel (such as the pictograms from the lower left section of Figure 1). Players did not have access to all gates to begin with. Each gate was gradually introduced as they progressed through the challenges. All the gates were unlocked over the course of the first 10 puzzles and all three could be combined and used together freely only from puzzle 12. They were able to place these under one of the available slots below the Qubit0 or Qubit1.

All the Quantum Odyssey puzzles used in this pilot consisted of building quantum circuits on two qubits that were initiated in their ground state 00. This information is shown by the fact the balls always spawn on the Computation Map from the 00 bitstring and keep falling until they reach the bottom of the circuit to then spawn again unless the player finds a solution to solve the puzzle. The solution is depicted by a set of one or more static balls for each puzzle. For puzzle 11, we can see the description of the puzzle solution given by the two blue balls behind bitstring 00 and 11 in snapshots 2, 3 and 4 from Figure 1. The puzzles were built with the following Quantum Information Science (QIS) topics in mind, in groups of 3 puzzles per topic: puzzles one to three were on the topic of understanding how to flip two-qubit arrays; puzzles four to six required the player to place qubits in superposition; and puzzles seven to nine required the player to use quantum interference to revert the qubits to their ground state (in other words the 00 state) from an initial superposition state. Puzzle 10 then challenged the player to figure out how to construct an entangled state on their own by simply being given the desired outcome in bitstrings (00 and 11), while the solution for puzzle 11 was to extract the same entangled state from a maximally entangled quantum state of equal amplitudes but a phase difference of -1 on bitstring 11, which was prepared by scrambling the entangled state with a H gate. Puzzles 12 to 21 were then built from combinations of two or more of these concepts.

Figure 1 presents a series of snapshots showing the most optimal way a player could solve puzzle 11 in Quantum Odyssey. Snapshot 1 shows the Circuit Configuration panel, where players can place quantum gates. The players were presented with a hidden maximally entangled quantum state that was scrambled with a Hadamard operator (H gate) on the Qubit 1. The challenge is encoded in the custom gate visible in snapshot 1 with a question mark symbol and the effect of the custom gate on the circuit is given to the player in the Computation Map (snapshot 2). Players had two types of gates available that can be dragged to the Circuit Configuration panel to one of the open slots, shown at the bottom of snapshot 1. The Computation Map (snapshots 2, 3 and 4) shows in real time the effect of all the gates placed in the Circuit Configuration. Players had to closely inspect the map and recognize that they could use what they learned in the previous puzzles: that they need to recover a maximally entangled quantum state that was scrambled in order to efficiently solve the puzzle, by just adding a single H gate in order to achieve quantum interference (snapshot 3) such that opposite phases collide on bitstrings 01 and 10 and phases of similar color enhance the ball sizes on 00 and 11. Thus, the effect of quantum interference is visually indicated with particle effects in snapshot 4 and the player obtains balls of the same size and color as indicated by the desired puzzle solution. The correct solution was to drag the gate labelled H to the slot on the right, as indicated by the white line in figure 1.

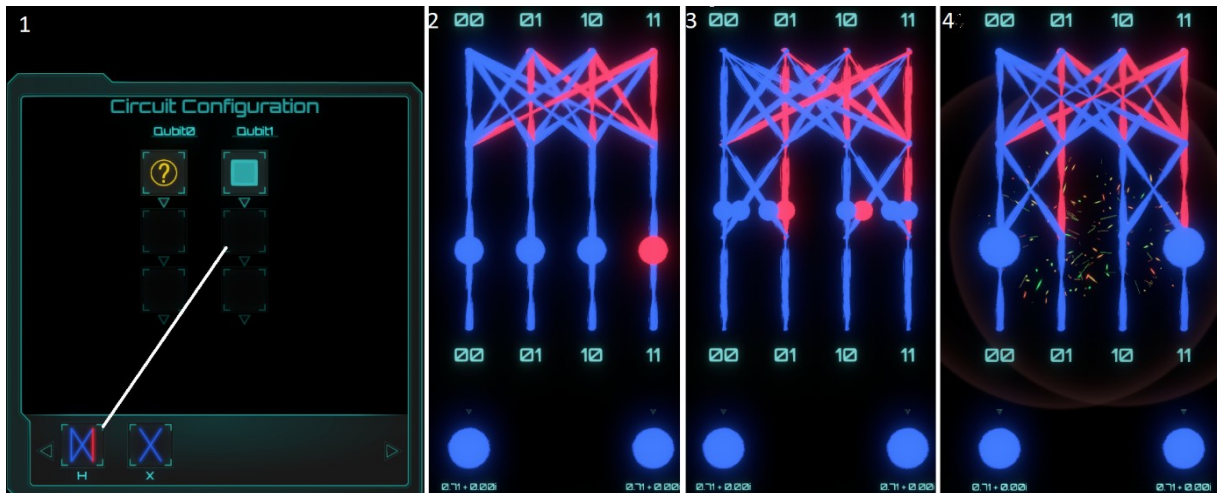

*Figure 1: Snapshots of puzzle 11 in the Quantum Odyssey puzzle game.*

**Research methods**

The project sought to answer the following research questions: Can quantum computing/quantum dynamics successfully be taught to non-specialists with a puzzle visualization tool? Can players with no background in quantum computing develop the skills needed to solve quantum compilation problems? Further questions were: How do different groups of players perform in the puzzle tasks? Does awareness of the puzzles being based on quantum computing affect performance? What is actually being learned? What interface/narrative framing best scaffolds learning for different user groups? Are there outstanding barriers to quantum literacy even without the need for advanced knowledge in Mathematics and Physics? Does a puzzle visualization tool increase motivation to engage in further learning about computational thinking and specifically quantum computation?

We worked with small groups of five students at a time, in two secondary schools in the UK, one from a disadvantaged catchment (involving separate groups from KS3, KS4, KS5 students and teachers), and one university (involving undergraduate students from different departments). In total, 10 groups took part in the project. Quantitative data were collected about use of Quantum Odyssey in the form of log files with raw data including saved puzzles, time stamps, and players' ratings of confidence at the end of each puzzle. Qualitative data were collected in the form of pre- and post-test questionnaires with both closed and open questions, observations and short focus group discussions with the workshop participants on learning, engagement and perceptions of the tool. The research design was mixed methods and cross-sectional, encompassing a quasi-experimental design, testing the underlying assumptions about learning through pre- and post-tests and analysis of log files. The exploratory part of the research design relied on inductive reasoning to interrogate the data through a thematic grounded theory analysis to derive concepts and factors important in the development of an appropriate learning model through learner questionnaires focused on navigation, screen-design elements, feedback on information provided, pedagogical components and focus group discussions on perceptions of the tool and the wider quantum computational concepts.

In order to facilitate engagement of participants, we drew on Gagne's Nine Events of Instructional Design (Gagné, Briggs, and Wager 1992) to structure the sessions. This identifies conditions for learning based on the information processing model of the mental events that occur when learners are presented with various stimuli. Sessions were structured to fit into one hour, with an initial pre-test questionnaire, playing the game for around 30 minutes, before completing the post-test questionnaires and taking part in a short focus group. We utilised this framework to trial different arrangements for scaffolding, from provision of only the visual puzzle game with no supporting information, to more scaffolding of the learning experience for participants. This model was then used to test performance and engagement by different groups.

Minimal prior information for participants consisted of the provision of: basic instructions about the game and its functions; direction to try and solve a puzzle in the simplest way possible; and having the Circuit Configuration panel, the Computational Map and start and end states described. More scaffolding included: the integration of more direct help and instruction with respect to game play; the provision of puzzle solutions; and providing instruction about underlying quantum mechanics concepts. Additional HE sessions were planned for two groups of STEM students, and further sessions were planned with industry, but the Covid-19 pandemic prevented these from being carried out. We therefore report on the findings from the sessions run with secondary school students along with the two sessions that took place with HE students only. To ensure that the trends we are reporting are statistically significant we use significance testing as described in the appendix.

**Findings**

The responses for the participants to the pre-questionnaire are provided in table 1 below. We have included all the information other than if English was the first language of participants. In the case of the school students, this was the case for all but three students, whereas for the HE students, half did not have English as their first language. Whilst this was factored into presentation of session content, level of English was good in all cases where it was not the first language and so did not have an impact on ability to engage with the sessions. Equally almost all the school students stated that they planned to do further study in a STEM subject (all but three), at the level following the one they were working at, so this information is not included below. Three of the HE students said that they were also planning to continue formal study in STEM, one at Masters level, and two through free online courses. The subject areas of the HE students were as follows: group 9 one undergraduate (UG) humanities, four postgraduate taught (PGT) course humanities/social science; group 10 all UG in social science. The questionnaire consisted of 10 questions. Responses to questions 3-7 and 10 were on a Likert scale where 1= not at all interested and 5 = very interested. Responses to 8-9 followed a similar scale from 1 = nothing, to 5 = I know a lot about it.

| Pre-session questions | School 1 groups | | | School 2 groups | | | | HE groups | |
|---|---|---|---|---|---|---|---|---|---|---|
| | 1 (n=5) | 2 (n=5) | 3 (n=5) | 4 (n=5) | 5 (n=5) | 6 (n=5) | 7 (n=5) | 8 (n=4) | 9 (n=5) | 10 (n=5) |
| 1. Age | 13/14 | 16/17 | 16-18 | 12/13 | 11-13 | 17/18 | 14/15 | 17/18 | Yr 1 UG/ PGT | Yr 1/ Yr 2 |
| 2. Gender (m/f) | 3/2 | 3/2 | 4/1 | 5/0 | 4/1 | 4/1 | 3/2 | 3/1 | 1/4 | 1/4 |
| 3. Interest in science | 4.4 | 4.4 | 1.8 | 3.0 | 4.6. | 4.4 | 4.0 | 3.0 | 3.8 | 3.6 |
| 4… physics | 3.2 | 4.2 | 2.2 | 3.8 | 4.8 | 4.0 | 4.2 | 4.2 | 2.6 | 2.6 |
| 5. ... computing | 4.4 | 3.4 | 4.4 | 5.0 | 4.8 | 2.8 | 4.2 | 3.5 | 1.0 | 3.2 |
| 6..computer games | 4.4 | 3.6 | 4.8 | 4.8 | 4.6 | 3.0 | 4.2 | 3.0 | 2.2 | 2.8 |
| 7… puzzle games | 4.2 | 3.4 | 3.0 | 4.2 | 4.2 | 3.6 | 4.0 | 3.5 | 3.8 | 3.4 |
| | | | | | | | | | | |
| Knowledge of 8… quantum | 2.4 | 2.8 | 2.0 | 0.5 | 2.0 | 2.8 | 2.6 | 3.2 | 2.0 | 2.6 |
| mechanics | | | | | | | | | | |
| 9… quantum computing | 2.2 | 2.4 | 1.0 | | 0.5 | 1.8 | 1.8 | 2.6 | 2.0 | 1.5 | 1.6 |
| 10. Interest in quantum computing | 4.8 | 4.2 | 4.2 | | 4.2 | 4.6 | 4.2 | 4.4 | 4.2 | 3.6 | 3.8 |

Table 1: Pre-session questionnaire data. Responses to q3-7 and q10 were on a Likert scale where 1= not at all interested and 5 = very interested with q8-9 from 1 = nothing, to 5 = I know a lot about it.

The post-session questionnaires asked 12 questions, with a four closed questions on a Likert scale (1-4, where 1 = not at all and 5 = very much) and a eight open questions for which participants were asked to complete the sentences, as in 5-12 below. Open question responses were analysed inductively using descriptive thematic analysis (Braun and Clarke, 2008) and table 2 details all codes found through in vivo coding, that is adhering to the language of the participants.

| Post-session questions | School 1 (SJ) group | | | | School 2 (RGS) groups | | | | HE groups | |
|---|---|---|---|---|---|---|---|---|---|---|
| | 1 (n=5) | 2 (n=5) | 3 (n=5) | 4 (n=5) | 5 (n=5) | 6 (n=5) | 7 (n=5) | 8 (n=4) | 9 (n=5) | 10 (n=5) |
| 1.Ease of navigation | 4.6 | 3.8 | 4.0 | 4.2 | 4.2 | 4.6 | 3.8 | 4.5 | 3.6 | 3.4 |
| 2. Appeal of screen designs | 4.0 | 4.0 | 4.2 | 3.2 | 4.0 | 3.6 | 3.8 | 4.0 | 4.2 | 4.2 |
| 3.Enjoyment in playing puzzles | 4.6 | 4.2 | 4.0 | 4.0 | 4.8 | 3.8 | 3.8 | 4.2 | 3.6 | 4.4 |

| 4. Interest to find out more | 4.8 | 4.6 | 4.4 | N/A | 5.0 | 3.8 | 4.2 | N/A | 3.5 | 3.8 |
|---|---|---|---|---|---|---|---|---|---|---|
| | | | | | | | | | | |
| 5. I became discouraged when… | Couldn't figure out the puzzles; tried combination that didn't work; other people were ahead of me; tools didn't do what I expected; couldn't do it straight away; tried and I wasn't improving; stuck on a level; didn't explain the purpose; too much trial and error required; multiple lines of circuit to be filled in; what operations were doing in relation to each other; couldn't remember if already tried a configuration. | | | | | | | | | |
| 6. I was motivated when… | Got something quickly; succeeded on a puzzle/level; noticed a link/pattern; worked out more complicated puzzles; worked on first try; discovered significance of controls/function of variables; used similar methods on higher levels; got something right and understood why. | | | | | | | | | |
| 7. I can see the link between… | Function of H/Control X/different symbols etc; game and how quantum computing works; how entanglement and annihilation can be used; controls and the path a ball takes; visual trial and error and effect on understanding. | | | | | | | | | |
| 8. I can understand… | Functions/controls; what I am meant to do; how quantum computing works a bit; qubits and entanglement; colours; some of the paths/ball patterns; the design; the ideas/the idea behind the game; the methods required to create a desired output; the beauty in the simplest solution; game mechanics; different combinations and what they do. | | | | | | | | | |
| 9. The puzzles challenged my understanding of… | Functions of certain features; classical computing; puzzles; logic; binary; complex circuits; tasks and time; memory; looking at things from a different point of view; patterns and manipulation; computational thinking; problem solving; ordering of the components and visualisation effects. | | | | | | | | | |
| 10. I am interested in knowing more about… | Subject as a whole; quantum computing; how the game works; language used; what each piece means; controls used; physics behind it; quantum computing games. | | | | | | | | | |
| 11. The best part of playing the puzzle games is… | Understanding what the features do and implementing that; figuring out puzzles/solutions/colours/controls; completing levels; winning; rising in difficulty; working through it; that it's satisfying whilst challenging; understanding it's showing visual representations; gaining an understanding. | | | | | | | | | |
| 12. I felt I learned most when… | Spotted a new link/how to use a function; got well done/right answer; my attempts failed; struggled with a puzzle then found solution; got the hints leading to more progress; became less about trial and error; the theory was explained; reused certain combinations to solve different puzzles; could use | | | | | | | | | |

| | features correctly first time; my mistakes were explained; paused and thought about what was needed; discussed with others. |
|---|---|

Table 2: Post-session questionnaires. Q1-4 used a Likert scale where 1 = not at all and 5 = very much. Q5-12 were open responses questions. N/A indicates participants gave incomplete responses.

Questionnaire data show that groups had a relatively high interest in science overall pre-test, with an average across the groups of 3.7 on the 5-point Lickert scale and 3.6 for interest in physics. This was similar for computing at 3.7 (but with considerable variation between the groups), 3.7 for computer games and 3.7 for puzzle games. Pre-test the knowledge-base was low, with an average of 2.3 across the groups for knowledge of quantum mechanics and 1.7 for quantum computing, whereas interest in quantum computing was high, as we would expect from self-selecting participants and perhaps which we could speculate to be the case more broadly for students interested in STEM subjects and/or computer games, averaging at 4.2 across the groups. Interest to find out more, in the post-test questionnaire, had gone up slightly to an average of 4.3 across the groups. Enjoyment in playing Quantum Odyssey in and of itself, as a game, was high, averaging 4.1 across the groups, with 3.9 for the appeal of the designs and 4.0 for ease of navigation, therefore the game itself was clearly popular.

In addition to use of pre- and post-test questionnaires, qualitative data was also generated in short focus groups at the end of each session (lasting no longer than 10 minutes). In addition, participant comments and questions were captured in audio recordings of the sessions and through the use of observation notes. The data were also used to detail when and how sessions were scaffolded by the game founder and creator, who led the sessions.

Analysis of the comments from participants in the school focus group (groups 1-8) highlighted several themes. Firstly, that there was more trial and error earlier on in playing the game than was evident based on just observation alone. Secondly, participants felt hints and explanations were needed at earlier stages, that is before trial and error, rather than after. It was also felt that having the puzzles visibly displayed whilst any explanations were provided was important. This focus on trial and error led participants to comment on the fact that it felt more like a memory task at times and that as the levels increased in complexity there was so much to focus on at once, that the relationships in the earlier stages became harder to visualise; 'I couldn't remember if I'd already tried the configuration'. Students commented on the fact that they would have appreciated narrative commentary to support them; 'it would be good to have a storyline', and also scaffolding to reinforce what had been learned; 'No info appeared once completed task explaining why'…'this could sometimes cause the level to complete before I had a full understanding/intuition for using the new piece'. Comments also focused on the positive rewards of completing puzzles and the challenge in this; 'Even though we don't know what we're doing this is a nice way of building up a kind of a logic'…'It was addictive, when you figure out a puzzle, you feel a relief and then you want to go'…'It's just like finding and learning patterns'.

Focus groups with the older A-level, and HE participants discussed the fact that they did not see the links to the real world of quantum mechanics and wanted to understand the theory behind the abstract visualisations of the puzzles and key concepts at the outset with an A-level student stating; 'I did not understand the relation to physics and did not feel like learning physics …' The game was fun, yet the link between quantum physics and the puzzles wasn't as apparent'. The participants in these focus groups were not given any explanation behind the physics of the game prior to completing the questionnaires. Participants also commented that there should be more support available when someone was stuck on a puzzle, so the idea of hints came into subsequent sessions, to help when participants were stuck, as well as to reinforce functions that were learnt. Despite some concerns about what was being learnt, there was nonetheless a palpable underlying appreciation of the manner of learning as this excerpt from an undergraduate student in group 9 demonstrates:

> 'During the game, it made me a lot more interested in it. In school, I was never interested, and physics was my least favourite subject in school but here I was interested in knowing what was happening after I finished. I like trying something out rather than being spoken at. It is more interesting when you do something that uses it because then I can see it and [I am] more interested in listening – I've done this and I want to know why and now I want an explanation, but in GCSE, they just talk and you ask 'why' and they say you don't need to know that'.

Focus group discussions therefore revolved around two key issues with respect to learning; learning how (what might be the shortest route to solving a puzzle making best use of the functions for instance) and learning what (the link between the abstract visualisations and the theory), which we discuss below. Participants also discussed being interested in future real world applications of quantum computing.

The game consisted of 21 individual puzzles, and across all sessions only one of the participants was able to complete all 21. However, all participants were able to complete at least the first 11 puzzles within 30 minutes. The puzzles within Quantum Odyssey came without any additional explanations, so each player had to follow their own judgement and interpret what they saw in their own way, to solve the puzzles. Figure 2 (top) shows that the number of participants who completed a given puzzle decreased slowly until puzzle 16, at which point the number decreased more rapidly.

Given that video games have traditionally been marketed more strongly toward a male audience, it is important to consider gender bias during the development process to ensure that questions perform equally well for all users. This has been explored below by considering the ratio of puzzles completed by male and female participants (Figure 2 bottom). The data show that the fraction of female participants completing puzzles compared to male participants increases above puzzle 16, only decreasing for puzzle 21, which was only completed by a single male participant.

Puzzle 17 was the question which closest to half of participants completed (n=24 completed, 18 did not complete). Statistical significance testing was carried out to determine whether there was a difference between the gender of participants that completed puzzle 17 (n=8 female participants out of 12 completing puzzle 17 and n=16 out of 28 male participants). A significance of $p=0.077$ was calculated, which indicates that the difference between genders was not statistically significant (although it should be noted that the sample size was small).

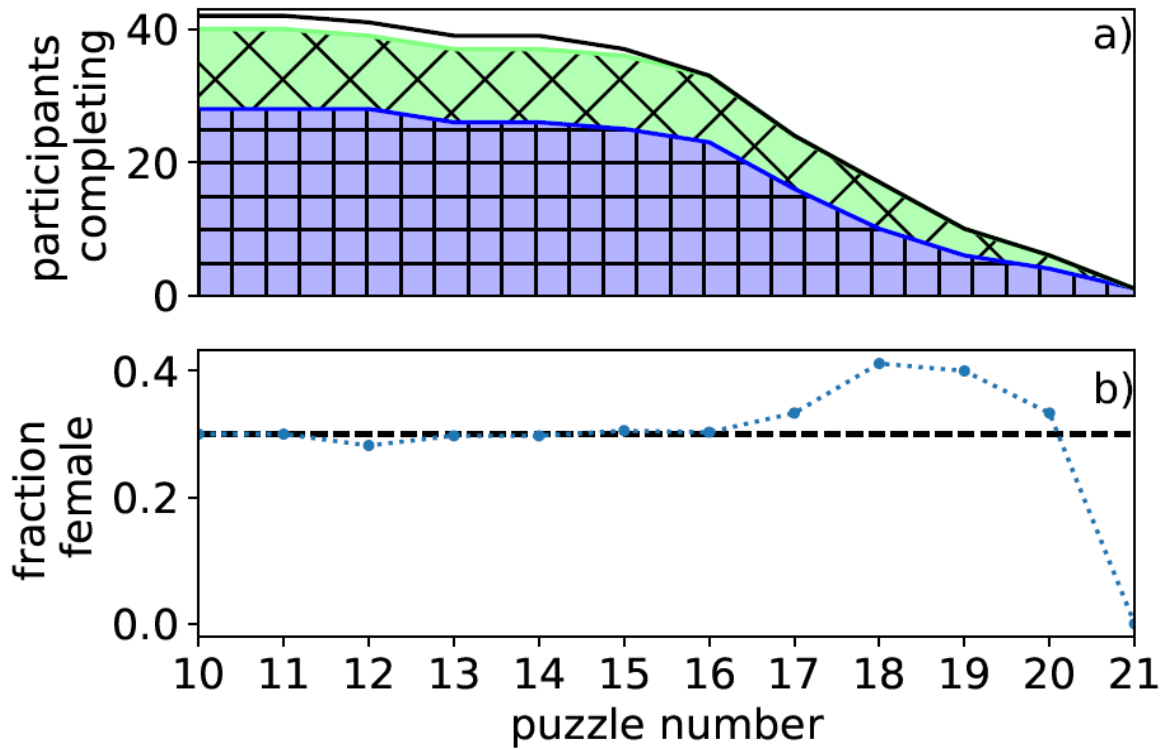

*Figure 2: Top frame: Number of participants completing a given puzzle, green colouring (above) indicates female participants, blue (below) indicates male, while uncoloured indicates participants who did not specify male or female as a gender. Bottom: Fraction of those who did specify male or female who specified female and completed different stages of the trial.*

Next, we examine whether there was a correlation between questionnaire responses about knowledge and interest and the highest puzzle solved. Figure 3 (top) shows the different levels of puzzle game interest for those completing different puzzle numbers. We see that more than half of the participants showed an interest of 4 or 5. The numbers drop off faster for those showing lower interest in these games. By puzzle 20, no participants with an interest level less than 3. The one participant who was able to complete all of the puzzles expressed an interest level of 5.

Figure 3 (bottom) shows the interest scores divided into histograms for those who did or did not complete puzzle 17. We find that interest in puzzle games for those who do not complete puzzle 17 is roughly evenly split between 1,2,3 and 4,5, but of those who do complete the puzzle more than three times as many indicated an interest of 4 or 5. A statistically significant difference ($p=0.019$) was found in completion rates for puzzle 17 between participants that selected 4 or 5 in response to the pre-session questionnaire relating to their "interest in puzzle games" and those that selected 1-3.

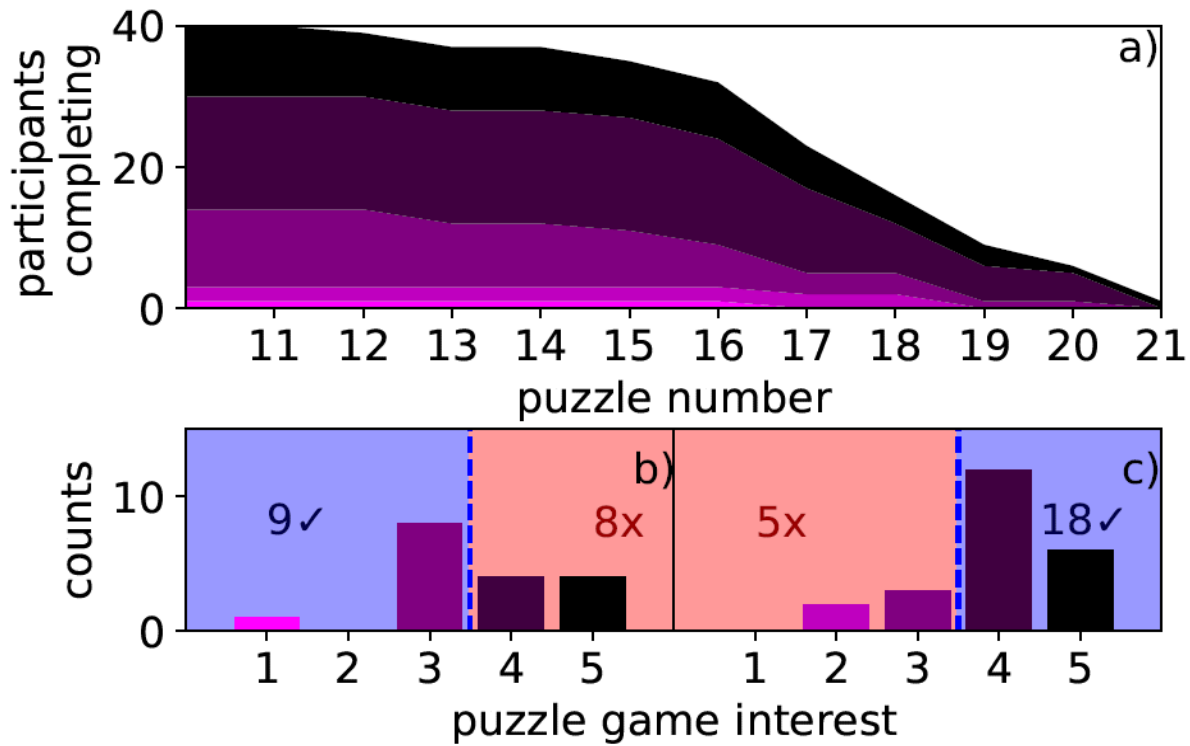

*Figure 3: Top: Number of participants completing different puzzles colour coded (lowest value on bottom) by interest in puzzle games. Participants who did not state their level of interest are excluded. Bottom left: Histogram of puzzle game interest for those who did not complete puzzle 17 bottom right: same as left but for those who did. Numbers give total counts in shaded areas (blue indicating successful prediction, red indicating failed).*

Next, we examine the performance of the participants on the first 11 puzzles, which all participants were able to complete. We again divide the participants into two groups, those who did complete puzzle 17 and those who did not. Figure 4 (top) shows the number of gates placed and Figure 4 (bottom) the completion time of each of these puzzles. We find that for many of the puzzles there is little difference between the two groups, especially in number of gates placed. However, those who completed 17 did solve puzzle 5-7 moderately faster, and placed fewer gates on puzzle 5. Interestingly, the group who completed puzzle 17 actually took longer and placed more gates on puzzle 11. Significance testing shows that the only statistically significant difference is for puzzle 11, indicating that participants who spend longer on puzzle 11 were more likely to complete puzzle 17. This effect achieves a significance of p=0.0040, indicating that it is very unlikely to be due to statistical fluctuations.

Based on the counter-intuitive finding that participants that spent a longer time to complete puzzle 11 were more successful in completing puzzle 17, it is worth examining a scatterplot of the time to complete puzzle 11 versus the highest puzzle solved by participants (Figure 5). The scatterplot is divided into four quadrants, with the dividing line on the y axis indicating the average time for completion of puzzle 11 and the division on the x axis showing the point at which half of participants managed to complete the puzzle (i.e. the point at which half of participants successful completed question 17). The four quadrants on the figure therefore show participants that took a) an above average time to complete puzzle 11 but did not complete puzzle 17 (n=5) b) a below average time to complete puzzle 11 but did not complete puzzle 17 (n=13) c) above average time to complete puzzle 11 and completed puzzle 17 (n=17) and d) below average time to complete puzzle 11 and completed puzzle 17 (n=7). As can be seen, of those that did not complete puzzle 17 (a and b), 2.6 times as many participants took less than the average time to complete puzzle 11 compared to above average time. For those who did complete puzzle 11 (c and d), 2.4 times as many participants took more than the average time, compared to below average time. Moreover, we see that many of those who took longer on puzzle 17 took much longer than the others, the longest time to complete was over 10 times as long as the shortest. The person who took the longest on this puzzle also was able to do well in terms of completing the later puzzles, completing puzzle 19, which only 6 participants out of 42 managed.

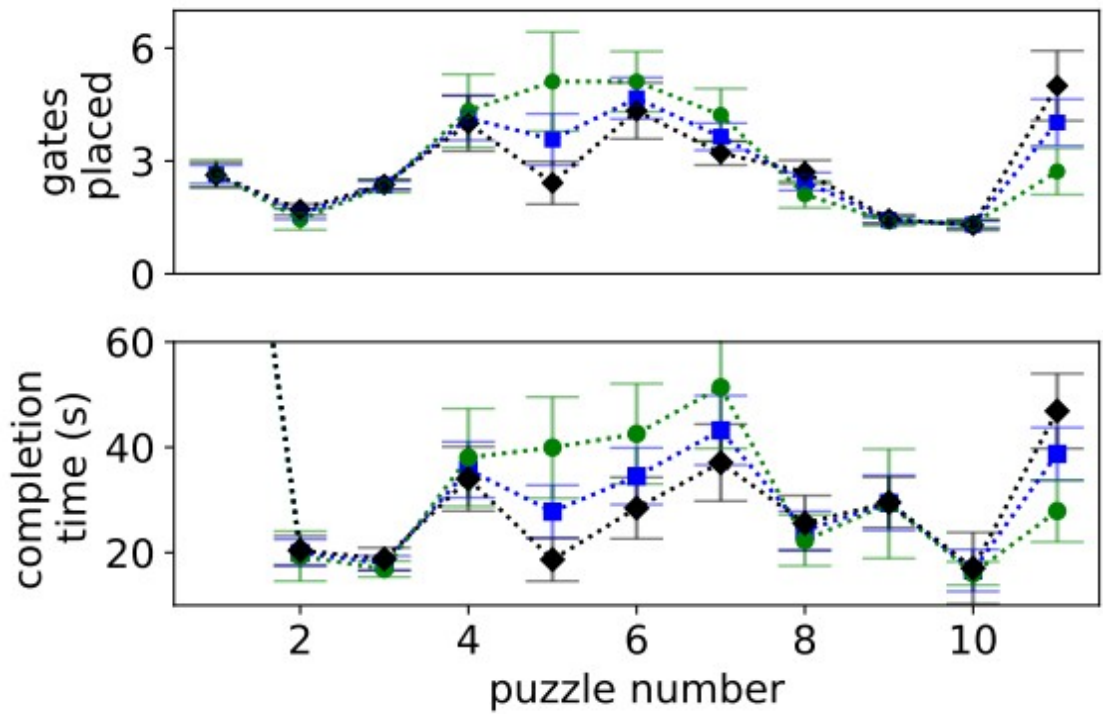

*Figure 4: Top: Number of gates placed for puzzles 1-11 (n=42). Bottom: completion time for the same puzzles (n=42). On both plots green circles indicate average values from those who did not complete puzzle 17, black diamonds indicate those who did, and blue squares show the average of all participants. Completion time of puzzle 1 is not shown because it was much larger for all groups, likely due to learning the controls.*

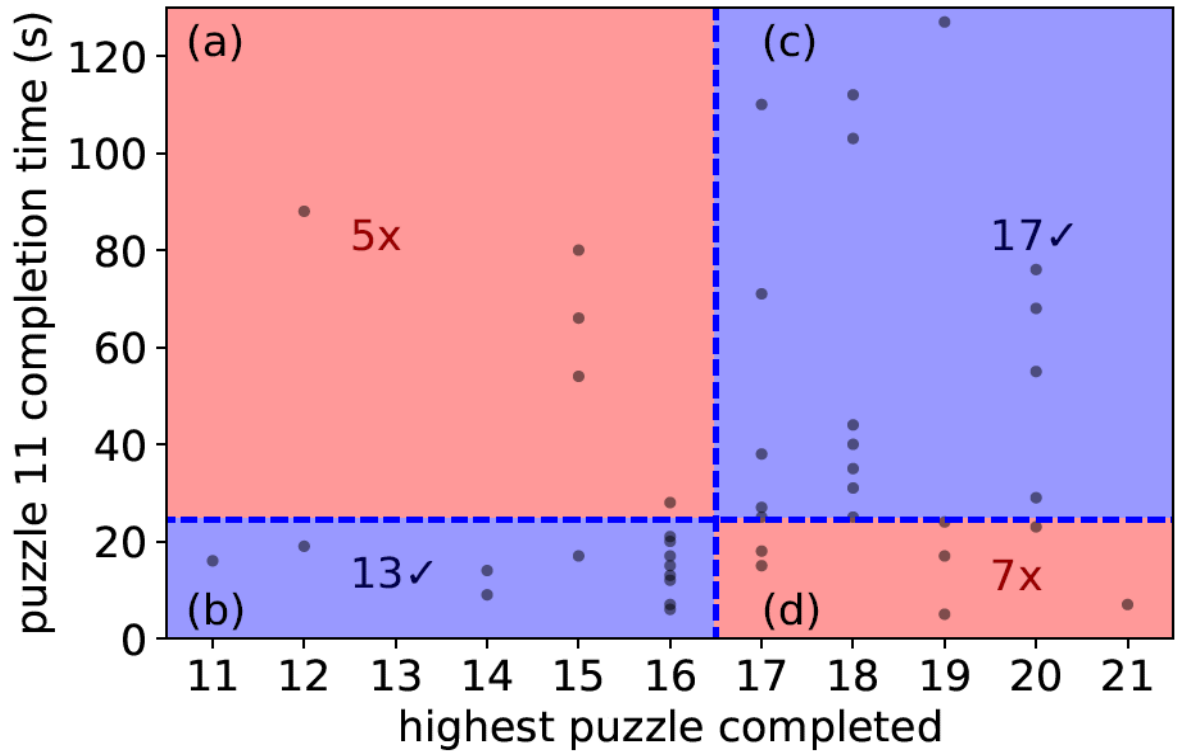

*Figure 5: Completion time for puzzle 11 versus the highest puzzle completed, lines show the most even possible division along each axis with colour coding to indicate regions corresponding to a successful versus unsuccessful prediction (green for the former, red for the latter). Counts in each region are also included for ease of reading.*

**Discussion**

These data provide preliminary evidence that Quantum Odyssey is potentially a powerful learning tool that facilitates game-based learning, with many elements of gamification incorporated in it. The project demonstrates that participants want to see further gamification elements introduced, such as a scoreboard and positive rewards for success, which fosters competition both on an individual level but also potentially between players. The gamification element was therefore important in terms of motivation and also an interest in puzzle games was predictive of success. We would therefore conclude from this that the gamified elements are a positive aspect of both motivation and success in moving through the game.

We found that the participants were discouraged when they felt they were struggling and in addition relying on trial and error rather than understanding the patterns and rules. They were also discouraged where they saw no purpose to what they were doing and motivated by understanding and applying this as they went through the puzzles. They were challenged by the differences that quantum computation posed as opposed to the logic of the classical or binary computational knowledge they held. They were also challenged by the need to memorise their learning about how to solve the puzzles as the difficulty grew. Of interest was the fact that responses to when most puzzles had been learned varied considerably, from when patterns were spotted and puzzles completed, to failing at puzzles, to working it out for themselves, to being given hints and having the theory explained. The point here is that this divergence of opinion in response to this question, as opposed to the more convergent range of opinions displayed in response to the other questions, shows that participants are not necessarily aware of how they best learn, or of what might facilitate this.

What was clear from the project's findings was a very strong desire from participants to have more scaffolding. The hints provided when players were stuck were appreciated as a way of mitigating too much trial and error play. Explanation of the gameplay was appreciated alongside the visuals in the game. Above all participants wanted links to theory and real-world applications to be made. As result of this a complete narrative text was written and recorded for all the puzzles and incorporated into the game after the completion of this project. The aim of this, based on the project findings, is threefold:

1. to reinforce learning through supporting memorisation of rules, detracting from the need for trial and error play and reinforcing core concepts learned;
2. to link these core concepts to the wider quantum mechanics theory and application, so as to situate the learning;
3. to add game-play elements to generate more control for the player, such as the option to replay narration and refer to both a bespoke encyclopaedia of key terminology and the game-play rules at will.

Qualitative data suggest that all these elements would increase both intrinsic motivation and have learning gains. As we found that interest in puzzle games was predictive of success in playing, we used the data to design further game-play elements that were suggestive of increasing the game's appeal and therefore user-engagement. Since we found that the *longer* players spent on earlier puzzles was predictive of later success, we subsequently designed a narrative storyline that would ensure engagement with earlier puzzles and so that rules learned were more likely to be achieved. The important consideration was to prevent a player moving through the puzzles too quickly, through too much trial and error play, as this would not sustain either motivation of level of success.

In terms of the research questions posed we found that the players of Quantum Odyssey were able to solve quantum computation problems even if they had never been exposed to quantum computation before. In addition, with appropriate scaffolding, what is being learned is therefore the ability to intuitively work with established QIS topics without having any previous exposure to the underlying mathematics or other sources that describe these topics. During the focus groups participants were asked technical questions about how to reach certain quantum states without having the in-game Computation Map in front of them and as a group were able to correctly understand the question and provide working answers, by recalling what quantum gates were doing and how they combined to build quantum algorithms. We can surmise that given the visuals and game-play experience were rated as highly enjoyable, and the visuals are an accurate and concrete representation of the underlying abstract quantum states, Quantum Odyssey functions to support the active participation of the learner and demonstrates what Karakostas (2004) refers to as an example of 'active scientific realism'. A powerful visualization tool such as this, to support the learning of complex QIS phenomena, can therefore help students to manage and engage with uncertainty.

**Conclusion and subsequent developments**

With the developments that have been implemented to the game as a result of these findings, we suggest that Quantum Odyssey can become a highly effective curricular model in the visual mode (Gilbert, 2005). In particular, the process of managing the students' uncertainty over both the intrinsic methodological holism of quantum computation and their epistemic shift from classical to quantum mechanics knowledge, can be accommodated through a narrative scaffolding that has been designed in response to these findings (Chen 2021; Roth et al. 2011). It is clear that the overlaying narrative that was needed is one that links puzzle patterns to realist knowledge in quantum mechanics and to real-world applications, ideally so as to situate the knowledge as much as possible. Given that many of our participants were not highly advanced in STEM, this linkage is an important finding from the data. The narrative creates a scaffold to structure and support the learning, based on the data collected in this project, with more rapid feedback built into the game for players on how they are progressing. This more scaffolded version of Quantum Odyssey will support more systematic analysis of learning in subsequent research.

**Funding information**

LN, HC, LMS, and NC were funded by UKRI (United Kingdom Research and Innovation) grant number BB/T018666/1. LN was supported by a Durham University PhD studentship, GD was supported by a Turkish government PhD studentship, and NC was also supported by UK Engineering and Physical Sciences Research Council Grant number EP/S00114X/1.

**Appendix: significance testing**

We adopt a simple statistical significance test for analysing whether different variables influence each other in a statistically significant way. To do this, we divide the data as evenly as possible based on each of the variables and then calculate how many samples are in the upper or lower half of both distributions versus being in the lower of one and the upper of the other. We consider the former to be cases where the variables successfully predict each other and the latter where they fail to predict each other. To calculate the statistical significance of a prediction, we find the probability that a fair coin flip (uncorrelated 50% probability of being in the top or bottom half of each distribution) could produce at least as large a fraction of successful predictions. The significance can be calculated using the following formula:

$$p = \frac{1}{2^{n_{fail}+n_{success}}} \sum_{i=0}^{n_{fail}} \binom{n_{fail}+n_{success}}{i}$$

where is the number of times the variables failed to predict each other, is the number of times the prediction was successful, and is the number of possible ways to choose k objects from a set of n. We note that this type of analysis is only valid when the samples can be divided roughly evenly with respect to at least one of the variables, fortunately, this is true for the cases we consider here. By convention a value of is considered statistically significant, while a higher value is not. By symmetry a value of indicates that the variables predict each other, but are anti-correlated, in other words a high value of one variable predicts a low value of the other and vice versa.

**Acknowledgements**

The authors would like to thank Rohit Sharma for useful discussions and for assisting in some data collection.